# Non-static ground state in SmB$_6$: a form of time crystal?

Qi Wu[1] and Liling Sun[1,2,3]

*[1]Institute of Physics, Chinese Academy of Sciences, Beijing 100190, China*
*[2]University of Chinese Academy of Sciences, Beijing 100190, China*
*[3]Collaborative Innovation Center of Quantum Matter, Beijing 100190, China*

**Here we propose the existence of a non-static ground state in the Kondo insulator SmB$_6$, with a unique accompany-type valence fluctuation of Sm ions in its bulk. Whether SmB$_6$ is a fashion of time crystal is an intriguing issue.**

Time crystal, a theoretical concept proposed by Wilczek[1,2] in 2012, has attracted a great deal of attention from scientists due to its unusual application background and historical reasons[3-5]. However, many proposed models for constructing a real time crystal have been denied [6-9], though some efforts are still being made[3]. The arguments on the time crystal mainly focus on whether there exists a real system with a non-static ground state. Here, we propose that such a state may exist in the Kondo insulator SmB$_6$.

SmB$_6$ is a mixed valence compound and has been one of the most puzzling materials for decades in the research field of strongly correlated electron materials. It shows abundant low-temperature phenomena which are not fully understood, such as resistance plateau[10], mixed valence state[11], anomalous quantum oscillation under magnetic field[12,13]. Especially, there exists a metallic surface state coexisting with an insulating bulk [14,15]. Recently, we propose that the accompany-type valence fluctuation state (AVFS) of Sm ions existed in the ground state may play a vital role for the emergence of its low-temperature anomalies[16]. The AVFS can be described by the valence change of Sm ions in the way of Sm$^{2+}$(NM)$\leftrightarrow$Sm$^{3+}$(M)+$d$ (here NM stands for non-magnetic ions and M stands for magnetic ions), and the corresponding configuration change of outer shell electrons ($4f^6\leftrightarrow4f^5$+5$d^1$). On the other aspect, the AVFS also implies that the populations of the magnetic Sm ions ($4f^5$ electrons with magnetic moment) increase or decrease together with the $d$ electrons. The AVFS can be tuned either by pressure or temperature [17-19].

The schematic crystal structure and configuration of electrons of Sm ions in $SmB_6$ is shown in Fig. 1, illustrating the non-static ground state. It can be seen that the Sm ions (as indicated by orange spheres) with the accompany-type valence fluctuation lodge in a bulk insulating phase. This unique structure can be also described as that some Sm ions are isolated (or trapped) in the insulating quantum wells that are constructed by the $B_6$ frameworks, which constructs a special structure with an array of the valence-fluctuating Sm ions in the insulating bulk. It is noteworthy that only a small amount of Sm ions are involved in such valence fluctuation (as indicated by orange spheres in Fig.1b), while most of Sm ions remain in a static mixed-valence state (see dark pink spheres in Fig.1b). In the ground state, the outer shell electrons of the Sm ion are oscillating (or fluctuating) intrinsically between the two states of $4f^6 \leftrightarrow 4f^5 + 5d^1$.

The reported experimental results on a variety of intimately related to the valence fluctuations including charge, magnetism and valence in the ground state evidence the existence of such a non-static ground state[20-25].

Finally, we note that the non-static ground state in $SmB_6$ seems to match all the conditions to be a time crystal suggested by Wilczek: (i) the particles of the system move and return to their original state; (ii) no energy exchange with its environment; (iii) the system is in its ground state; (iv) the movement is derived by a special form of perpetual motion in which the continuous time translation symmetry spontaneously break[1,26-28] (in the AVFS the orbital change of $f \leftrightarrow d$ violates parity conservation[29]). Consequently, an intriguing question on whether the $SmB_6$ with such an array of non-static Sm ions in its insulating bulk belongs to an exotic fashion of time crystal is raised.

**Acknowledgements**

Authors thank W Yi for his assistance in the figure preparation. The work in China was supported by the National Key Research and Development Program of China (Grant No., 2016YFA0300300 and 2017YFA0302900), the NSF of China (Grants

No. 91321207, No. 11427805, No. U1532267), the Strategic Priority Research Program (B) of the Chinese Academy of Sciences (Grant No. XDB07020300).

Correspondence and requests for materials should be addressed to Q.W. (wq@iphy.ac.cn) and L.S. (llsun@iphy.ac.cn).

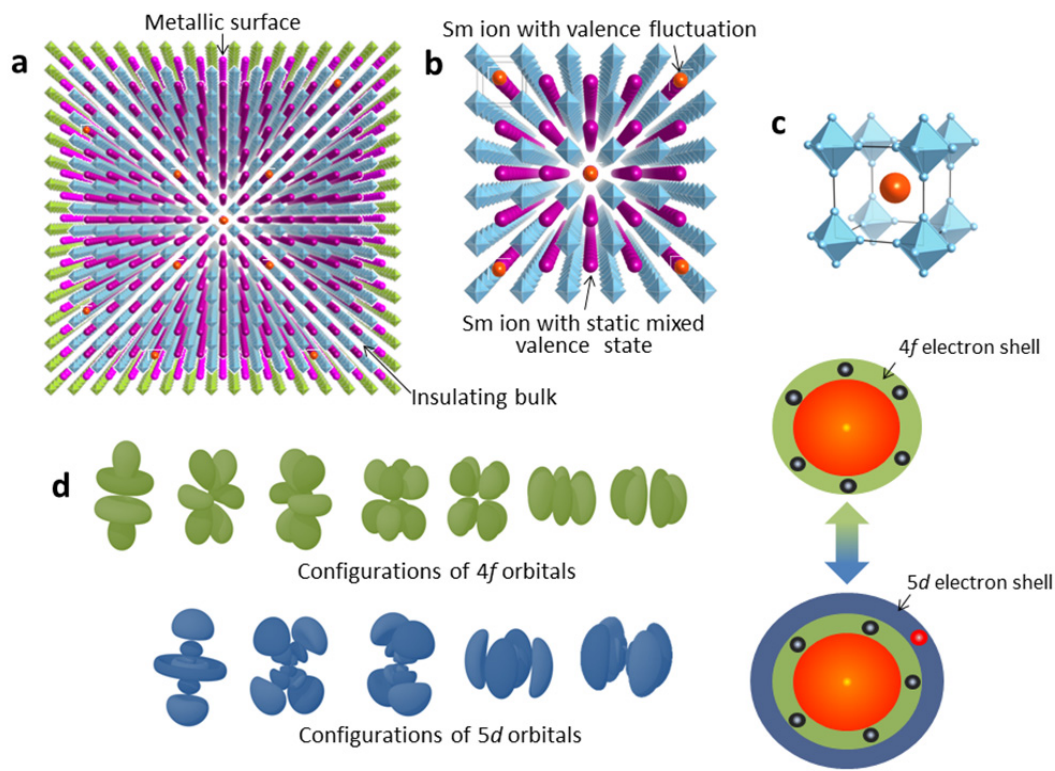

**Figure 1 Schematic crystal structure and configurations of electrons of Sm ions in SmB₆, illustrating a non-static ground state. (a)** The section view of the crystal structure of SmB₆. The green octahedrons (B₆ framework) and the light pink spheres (Sm ions) represent the structure of metallic surface. While the light blue octahedrons and the dark pink spheres stand for the insulating bulk structure. **(b)**

The enlarged view of the bulk structure, which is composed of $B_6$ octahedrons and Sm ions. There two types of Sm ions in the bulk structure: one is in a static mixed-valence state (as indicated by the dark pink spheres), and the other is in a non-static valence fluctuation state (as indicated by orange spheres). **(C)** The unit cell displays a Sm ion with valence fluctuation in the interstitial of $B_6$ framework. **(d)** The illustration of the accompany-type valence fluctuation of a Sm Ion. The green ring represents the $4f$ shell and the blue ring stands for the $5d$ shell. The seven green configurations represent the $4f$ orbitals and the five blue configurations represent the $5d$ orbitals. The black spheres stand for the $4f$ electrons and the red sphere stands for a $5d$ electron. The outer shell electrons of the Sm ion are oscillating between the two configurations of $4f^6 \leftrightarrow 4f^5 + 5d^1$.